\documentclass[cmfonts,numberedheadings,final]{aipproc}

\usepackage{times}
\layoutstyle{6x9}

\begin{document}

\title{New Circumstellar Dust Creation in V838 Monocerotis}


\author{John P. Wisniewski}{
 address={University of Washington}
  ,altaddress={NSF Astronomy \& Astrophysics Postdoctoral Fellow}
}

\author{Mark Clampin}{
  address={NASA GSFC}
}

\author{Karen S. Bjorkman}{
  address={University of Toledo}
}

\author{Richard K. Barry}{
  address={NASA GSFC}
}

\begin{abstract}
We report high spatial resolution 11.2 $\mu$m and 18.1 $\mu$m imaging of the eruptive variable 
V838 Monocerotis, obtained with Gemini Observatory's Michelle in 2007 March.  The 2007 flux density 
of the unresolved stellar core is roughly 2 times brighter than that observed in 2004.  We interpret these 
data as evidence that V838 Mon has experienced a new circumstellar dust creation event.  We also report 
a gap of spatially extended thermal emission over radial distances of 1860-93000 AU from the central 
source,  which suggests that no prior significant circumstellar dust production events have occurred within 
the past $\sim$900-1500 years.
\end{abstract}

\maketitle

\section{Introduction}

V838 Monocerotis experienced three dramatic photometric outbursts in 2002 January-March in which it 
brightened from v $\sim$15 to v $\sim$6.6.  The star displayed a F-type spectrum characterized by 
strong P-Cygni profiled H$\alpha$ emission shortly after outburst (see e.g. Wisniewski et al 2003a), but 
rapidly cooled to a L-type supergiant by 2002 October (Evans et al 2003).  A spectacular light emerged 
around the system in 2002 February, and the time evolution of scattered light from this echo has been 
traced in exquisite detail by the Hubble Space Telescope Advanced Camera for Surveys (see e.g. Bond et 
al. 2003).  Spatially resolved thermal emission from the light echo material was also detected by 
the Spitzer Space Telescope's MIPS (Banerjee et al. 2006).

The origin of the light echo material (circumstellar versus interstellar) is not well known, but the 2004 
epoch detection of the echo in thermal emission enabled Banerjee et al (2006) to place initial constraints 
on the dust mass of the material.  The large mass inferred led Banerjee et al (2006) to suggest that at 
least some of the dust might be interstellar in nature.  

\subsection{Key Questions}

While a multitude of research groups have observed and monitored V838 Mon over a wide-range of wavelengths and using 
a wide-variety of diagnostic techniques, several fundamental questions regarding the object remain unanswered.  These 
include: 

\begin{itemize}
\item \textit{What caused V838 Mon's outbursts?}  A diverse array of mechanisms have been proposed to explain V838 Mon's 
behavior, as summarized in the proceedings of the 2006 conference ``The Nature of V838 Mon and Its Light Echo'' (ASP Conf Proc Vol 324).  These mechanisms can be classified into two groups: events which are singular in nature (i.e. a stellar merger) and 
events which are recurrent (i.e. a nova-like explosion).

\item \textit{How much of V838 Mon's light echo material is remnant circumstellar material and how much is interstellar material?} 
\end{itemize}

\section{New Mid-IR Observations}

We imaged V838 Mon at N' (11.2 $\pm$ 2.4 $\mu$m) and Qa (18.1 $\pm$ 1.9 $\mu$m) on 2007 March 21-22 using 
Gemini Observatory's Michelle instrument.  The data were obtained during stable weather conditions, resulting in observed 
FWHM values of 0.40 $\pm$ 0.02 arcsec at N' and 0.60 $\pm$ 0.05 arcsec at Qa.  The Michelle FOV is 32 x 24 arcseconds.  
For comparison purposes, we also report the mid-IR flux density of V838 Mon from Spitzer MIPS observations at 
23.7 $\pm$4.7 $\mu$m and 71.42 $\pm$19 $\mu$m, obtained on 2007 April 10 as part of program 30472 (PI K. Su).  The 
FWHM of MIPS data at 24 and 70 $\mu$m is 6 and 18 arcseconds respectively.  Full details regarding the basic reduction 
and processing of these data can be found in Wisniewski et al (2008).

\section{Results}
\subsection{Mid-IR Flux Density of the Unresolved Stellar Core}
We find that the mid-IR flux density of the unresolved stellar core of V838 Mon increased by a factor of $\sim$2 from 2004 to 
2007 (see Figure 1).  The expected photospheric flux at these wavelengths is nominal; we interpret these data as clear evidence 
that V838 Mon has experienced a new circumstellar dust creation event.  We suggest that this dust has condensed from the 
expanding ejecta from the 2002 outburst events.  Such a scenario was foreshadowed by Lynch et al (2004), who reported 
the detection of an expanding region of molecular gas from the outburst events, which they noted was likely dust precursor 
material.

\begin{figure}
  \includegraphics[height=0.35\textheight]{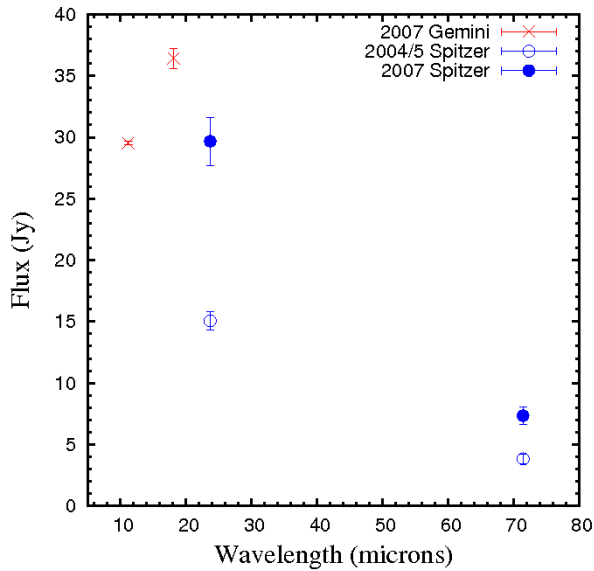}
  \caption{This figure, adopted from Wisniewski et al (2008), shows that the 2007 mid-IR flux density 
of the unresolved central core of V838 Mon is $\sim$2 times stronger than that observed in 
2004.  We interpret these data as clear evidence that V838 Mon has experienced a new 
circumstellar dust creation event.}
\end{figure}

\subsection{Spatially Resolved Thermal Emission from the V838 Mon Light Echo}

Our ground-based Gemini Michelle data have a ten-fold better spatial resolution as compared 
to analogous Spitzer MIPS data, and therefore enable us to search for evidence of mid-IR thermal emission from 
V838 Mon's light echo material in regions  
interior to that probed by Banerjee et al (2006).  After subtracting the stellar PSF from our Gemini data, we detect 
no evidence of extended thermal emission above a level of $\sim$1 mJy at 11.2 $\mu$m and $\sim$7 mJy at 18.1 $\mu$m 
over radial distances of 1860 AU (0.3 arcseconds) -93000 AU (15.0 arcseconds; see Figure 2).  Using the simple 
assumption that the 2002 ejecta material 
expands at a constant velocity of 300-500 km s$^{-1}$, this gap of spatially resolved thermal emission suggests that 
no significant prior circumstellar dust production events have occurred within the past 900-1500 years.  If the 
mechanism which produced V838 Mon's 2002 outbursts were the byproduct of a recurrent phenomenon (i.e. a nova-like 
explosion), our results suggest that the time-scale of such events are at least of order 10$^{3}$ years.

\begin{figure}
  \includegraphics[height=0.4\textheight]{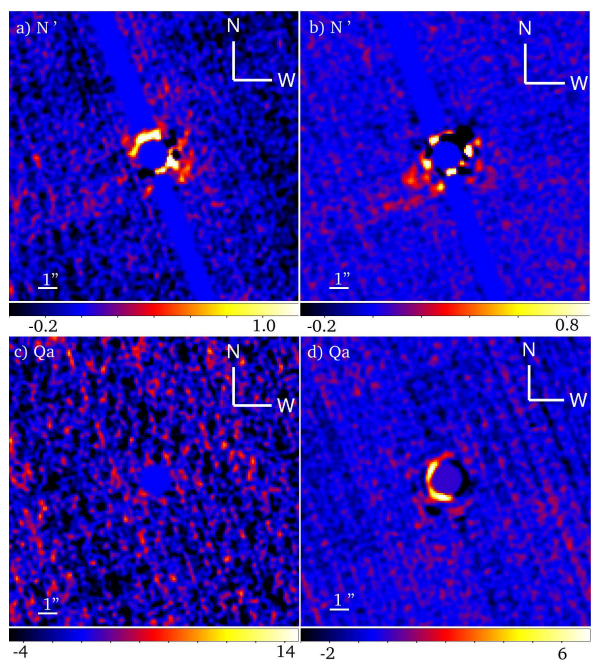}
  \caption{PSF-subtracted Gemini 11.2 $\mu$m (top panels) and 18.1 $\mu$m (bottom panels) imagery 
from 2007 March 21-22, as adopted from Wisniewski et al (2008).  The 15 x 15 arcsecond field of view lies inside of the 
extended mid-IR emission detected by Spitzer.  We detect no evidence of extended emission in our Gemini 
data above a level of $\sim$1 mJy at 11.2 $\mu$m and $\sim$7 mJy at 18.1 $\mu$m over radial 
distances of 1,860 AU (0.3 arcseconds) - 93,000 AU (15.0 arcsec).}
\end{figure}

\section{Future Observations}

Our observations have revealed clear evidence that V838 Mon has begun to produce new circumstellar dust.  We 
encourage continued mid-IR monitoring of the star to trace the evolution of this event and diagnose its chemistry.  
If this dust-forming ejecta has reached the location of V838 Mon's B3V secondary as suggested by Bond (2006), new 
interferometric observations might reveal detectable changes in the visibility of the system as compared to 2004-epoch 
interferometric observations by Lane et al (2005).

\begin{theacknowledgments}
  Our Gemini observations were obtained in program 2007A-Q-68.  JPW acknowledges past support from a 
NASA NPP Fellowship (NNH06CC03B)  and is currently supported by a NSF Astronomy \& 
Astrophysics Postdoctoral Fellowship (0802230).
\end{theacknowledgments}

\end{document}